\documentclass[11pt]{article}             
\usepackage{graphicx}
\begin{document}
\begin{center}
The sum rules for the spin dependent structure functions $g_1$ in
the isovector reaction\vspace*{5mm}\\
Susumu Koretune\\
Department of Physics,  Shimane University,\\
Matsue,Shimane,690-8504,JAPAN
\end{center}
\begin{abstract}In the isovector reaction, the sum rule 
for the spin dependent function $g_1$ which
is related to the cross section
of the photoproduction is derived. In the small $Q^2$ region, 
the sum rule is dominated by the low energy contribution and 
it tightly connects the resonance, the elastic, and
the non-resonant contributions.
\end{abstract}
11.55.Hx,12.38.Qk,13.60.Hb\\

It has been known that some sum rules derived from the canonical
quantization on the null-plane get the contribution from the nonlocal 
quantity corresponding to the matrix
element of the bilocal current which is absent in the equal-time 
formalism\cite{DJT}. The sum rule for the spin dependent
function $g_1^{ab}$ corresponding to the moment at $n=0$ is one
example,where $a,b$ denotes the flavor suffix of the currents. 
This sum rule is for the anti-symmetric combination with respect to
$a,b$. The corresponding sum rule in the equal-time formalism 
had been considered peculiar since it was invalid in the free
field model. This fact was discussed in
Ref.\cite{beg}, and also in Ref.\cite{Adler}. The null-plane method
circumvented this defect. 

Recently, there is a great experimental interest 
in the behavior of the polarized structure functions in the low $Q^2$
region\cite{Bur}. Motivated by this, the sum rule for the $g_1^{ab}$
derived from the connected hadronic matrix element of the 
current anti-commutation relation on the null-plane has been
transformed to the one which is sensitive to the behavior in this region.\cite{kore05}
Here, we report the same method can be applied to the 
sum rule known in the null-plane formalism based on the current
commutation relation, and transform it to the experimentally testable
form. It should be noted that the current commutation relation
on the null-plane is an operator relation while 
the current anti-commutation relation on the null-plane exists
only as a stable hadronic matrix element. We can derive the latter
from the former but we can not do the converse. Further, the sum rule derived
here is a nonsiglet quantity and that in Ref.\cite{kore05} include a
singlet quantity. This difference is reflected in the high energy
behavior, i.e.; the superconvergence relation in the derivation of
the sum rule. 

According to Ref.\cite{DJT}, we obtain
\begin{equation}
 \int_{0}^{1}\frac{dx}{x}g_1^{[ab]}(x,Q^2)=-\frac{1}{16} f_{abc}\int_{-\infty}^{\infty}
d\alpha [A_c^5(\alpha ,0)+\alpha \bar{A}_c^5(\alpha ,0)],
\end{equation}
where $A_c^{5\beta}(x|0)$ is the anti-symmetric bilocal current, and its
matrix element is defined as
\begin{equation}
 <p,s|A_c^{5\beta}(x|0)|p,s>_c=s^{\mu}A_c^5(p\cdot x,x^2)+p^{\mu}(x\cdot
 s)\bar{A}_c^5(p\cdot x,x^2)+x^{\mu}(x\cdot s)\tilde{A}_c^5(p\cdot x,x^2).
\end{equation}
Since the right-hand side of the sum rule is $Q^2$ independent, we obtain for the
anti-symmetric combination with respect to $a,b$
\begin{equation}
 \int_{0}^{1}\frac{dx}{x}g_1^{[ab]}(x,Q^2)=\int_{0}^{1}\frac{dx}{x}g_1^{[ab]}(x,Q^2_0).
\end{equation}
Now, we take $Q_0^2=0$ and use the relation 
\begin{equation}
 G_1^{ab}(\nu ,0)=-\frac{1}{8\pi^2\alpha_{em}}\{\sigma_{3/2}^{ab}(\nu ) - \sigma_{1/2}^{ab}(\nu )\}
=-\frac{1}{8\pi^2\alpha_{em}}\Delta\sigma^{ab}(\nu )
\end{equation}
 By setting $a=(1+i2)/\sqrt{2}, b=a^{\dagger}$, and separating out the
 elastic contribution, we obtain the sum rule which relates the
$g_1$ and the cross section of the photo-production in the isovector reactions.
 
Now the Regge theory predicts as 
$g_1^{[ab]}\sim \beta x^{-\alpha(0)}$ with $\alpha(0) \leq 0$,
and hence the sum rule is convergent.
However, the perturbative behavior like the DGLAP fit to the unmeasured small $x$
region has large ambiguity\cite{grsv} and the sum rule is
possibly divergent. 
The double logarithmic $(log (1/x))^2$ resummation 
give more singular behavior than the Regge
theory\cite{bade} and the sum rule (3) is also divergent.
Though, whether the sum rule diverges or not can not be judged
rigorously by these discussions, it is desirable to discuss the 
regularization of the sum rule and give it a physical meaning even 
when the sum rule is divergent. Now, the regularization of the 
divergent sum rule has been known to be done by the analytical
continuation from the nonforward direction\cite{analy}.
We first derive the finite sum rule in the small but sufficiently large $|t|$ region
by assuming the moving pole or cut. Then we subtract the singular pieces
which we meet as we go to the smaller $|t|$ from both hand-sides of the
sum rule by obtaining the condition for the coefficient of the singular
piece. After taking out all singular pieces we take the limit $|t|\to
0$. The sum rule obtained in this way can be transformed to the form
where the high energy behavior from both-hand sides of the sum rule
is subtracted away.  Practically, if the cancellation at high energy is
effective, since the condition is needed only in the high energy limit, 
we can consider the sum rule irrespective of the condition. 
The sum rule of this type can be obtained as follows.

The hadronic tensor is defined as
\begin{equation}
W^{\mu\nu}_{ab}|_{{\rm spin\ dependent}}=\frac{1}{4\pi m_N}\int
 d^4x\exp{(iq\cdot x)}\langle
 p,s|[J_a^{\mu}(x),J_b^{\nu}(0)]|p,s\rangle_c|_{{\rm spin\ 
 dependent}}  .\\
\end{equation}
Since we take $a=(1+i2)/\sqrt{2}, b=a^{\dagger}$ which means to take
$J_a^{\mu}$ as $J_{1+i2}^{\mu}/\sqrt{2}$ and the state
$|p\rangle$ as the proton, the Born term
is given as
\begin{equation}
W^{\mu\nu}_{ab}|_{{\rm Born}}=\frac{1}{4\pi m_N}\int
 d^4x\exp{(iq\cdot x)}\sum_{s^{\prime} ,n}\langle
 p,s|J_a^{\mu}(x)|n,s^{\prime}\rangle\langle n,s^{\prime}|J_b^{\nu}(0)]|p,s\rangle_c,
\end{equation}
where $n$ in the intermediate state specifies both the neutron and its
momentum and the $n$ in the sum means to take the momentum integral. Then we define
 \begin{equation}
\langle p,s|J_{1+i2}^{\mu}(0)|n,s^{\prime}\rangle =
\bar{u}_s(p)(\gamma^{\mu}g_V^{+}+\frac{1}{2}(p+n)^{\mu}f_V^{+})u_{s^{\prime}}(n).
\end{equation}
where the form factors $g_V^{+}$ and $f_V^{+}$ are related to
the usual Dirac and Pauli form factors or Sachs form factors as 
$g_V^{+}=F_1^{+}+F_2^{+}=G_M^{+}$ and
$m_Nf_V^{+}=-F_2^{+}=-(G_M^{+}-G_E^{+})/(1+Q^2/4m_N^2)$.
It should be noted that the $+$ component of the
form factor is connected to the difference between the form factor of
the proton and that of the neutron. This is because
$J_{1+i2}^{\mu}(0)=[J_{3}^{\mu}(0),I_{+}]=[J_{em}^{\mu}(0),I_{+}]$
since the hypercharge current commutes with $I_{+}$,
where $J_{em}^{\mu}(0)$ is the electromagnetic current and $I_{+}$ satisfies
$I_{+}|n\rangle =|p\rangle$ and $\langle p|I_{+}=\langle n|$.
Then it is straightforward to take
out the Born term contribution in the spin dependent function $g_1^{ab}$.
Now we take $\nu_{c}^{Q} =m_pE_Q$ where $E_Q$ is given as
$E_Q=E_c+Q^2/2m_p$ with $E_c$ being the cut off energy of the photon
in the laboratory frame. By separating out the Born term we
rewrite the regularized sum rule as
\begin{eqnarray}
B(Q^2)+K(E_c,Q^2)&=&\int_{E_0}^{E_Q}\frac{dE}{E}[2g_1^{1/2}(x,Q^2)-g_1^{3/2}(x,Q^2)]\\\nonumber
&+&\frac{m_p}{8\pi^2\alpha_{em}}\int_{E_0}^{E_c}dE[2\Delta\sigma^{1/2}-\Delta\sigma^{3/2}],
\end{eqnarray}
by using the isospin rotation as in the Cabibbo-Radicati sum rule\cite{cab},
where $B(Q^2)$ is given as
\begin{equation}
B(Q^2)=\frac{1}{4}\{(\mu_p-\mu_n)-\frac{1}{1+Q^2/4m_p^2}G_{M}^{+}(Q^2)
[G_{E}^{+}(Q^2)+\frac{Q^2}{4m_p^2}G_{M}^{+}(Q^2)]\},
\end{equation}
with
\begin{equation}
G_E^+(Q^2)=G_E^p(Q^2)-G_E^n(Q^2), \qquad
G_M^+(Q^2)=G_M^p(Q^2)-G_M^n(Q^2),
\end{equation}
and
\begin{equation}
K(E_c,Q^2)=-\int_{E_Q}^{\infty}\frac{dE}{E}[2g_1^{1/2}(x,Q^2)-g_1^{3/2}(x,Q^2)]
-\frac{m_p}{8\pi^2\alpha_{em}}\int_{E_c}^{\infty}dE[2\Delta\sigma^{1/2}-\Delta\sigma^{3/2}].
\end{equation}
Here, the suffix $1/2$ or $3/2$ in $g_1$ and $\Delta\sigma$ means
the quantity in the reaction $(isovector\ photon) + (proton) \to
(states\ of\ isospin\ I)$ where $I=1/2,3/2$. Then, $g_1(x,Q^2)$ in the virtual charged photon
reaction $(g_1^{ab}(x,Q^2)-g_1^{ba}(x,Q^2))$ is transformed to the quantities in the
real neutral isovector photon corresponding to the vector current $J^{\mu}_3$ 
as $(2g_1^{1/2}(x,Q^2)-g_1^{3/2}(x,Q^2))$  by a simple isotopic analysis. 
Similar fact applies to $\Delta\sigma$.
As discussed in \cite{kore05}, if we take $E_c=2$(GeV$^2$) and a small $Q^2$, the
contribution from $K(E_c,Q^2)$ is expected to be small and almost
negligible. We can expect the same kind of things happens also in this
case. The contributions from the Born terms $B(Q^2)$
can be estimated by using the standard dipole fit, where Galster
parameterization  is used for the $G_E^n$.\cite{Galster}  The 
resonance contributions on the right-hand side of the sum rule (8)
can be estimated by the parameters given in
\cite{Sim} if we neglect the isoscalar photon contribution. 
The results are given in the figure.  From it, 
we see that, to satisfy the sum rule, the difference of the
the non-resonant contribution between 
$\int_{E_0}^{E_Q}\frac{dE}{E}[2g_1^{1/2}(x,Q^2)-g_1^{3/2}(x,Q^2)]$ and
$-\frac{m_p}{8\pi^2\alpha_{em}}\int_{E_0}^{E_c}dE[2\Delta\sigma^{1/2}-\Delta\sigma^{3/2}]$
is negative in the very small $Q^2$
region and becomes positive above some value near $Q^2\sim
0.15$(GeV/c)$^2$. This sign change occurs in the region where the change
of the difference between the resonances becomes small while that
between the Born terms is rapid. 
\begin{figure}
\includegraphics[width=100mm]{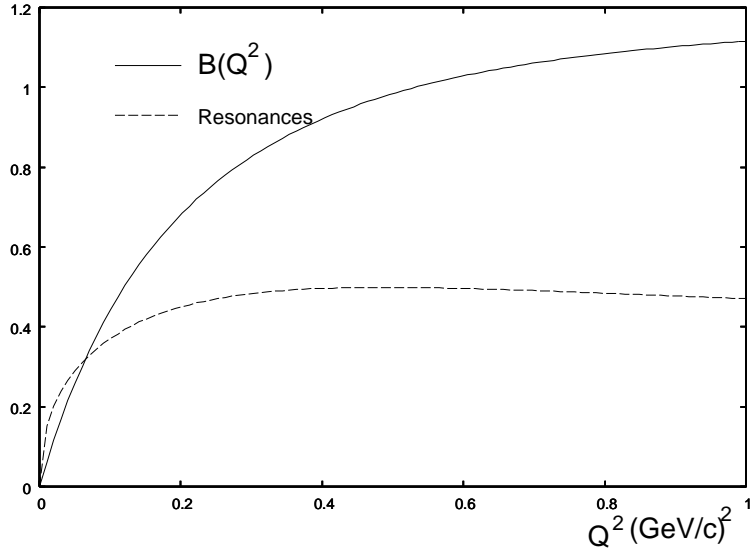}
\caption{The contributions from the Born terms as given by the $B(Q^2)$ and
those from the resonances.}
\end{figure}

In summary, in the isovector reaction, the sum rule 
for the spin dependent function $g_1^{ab}$ 
which is related to the cross section
of the photoproduction is given. By taking
the parameter in the sum rule appropriately, 
the sum rule is expected to be dominated by
the low energy contributions. Then, the sum rule shows that
the resonance, the elastic, and
the non-resonant contributions are
tightly connected in the small $Q^2$ region.

\end{document}